\documentclass[superscriptaddress,preprintnumbers,showpacs,prb,aps,twocolumn]{revtex4}

\usepackage{amsfonts}
\usepackage{amssymb}
\usepackage{amsmath}
\usepackage{graphicx}
\usepackage{natbib}
\usepackage{physics}
\usepackage{siunitx}
\usepackage{float}

\usepackage{fancyhdr}
\newcommand{\parallelsum}{\mathbin{\!/\mkern-5mu/\!}}

\setcitestyle{numbers,square}

\usepackage[usenames,dvipsnames]{color}
\usepackage{soul}

\usepackage{hyperref}

\begin{document}

\newcommand{\ie}{{\it i.e.}}
\newcommand{\eg}{{\it e.g.}}
\newcommand{\etal}{{\it et al.}}

\newcommand{\micron}{$\mu$m}
\newcommand{\smb}{SmB$_6$}
\newcommand{\kb}{k_{\rm B}}


\title{
{Field-dependent heat transport in the Kondo insulator SmB$_6$: \\ phonons scattered by magnetic impurities}}

\author{M-E.~Boulanger}
\affiliation{Institut quantique, D\'epartement de physique \& RQMP, Universit\'e de Sherbrooke, Sherbrooke, Qu\'ebec J1K 2R1, Canada}

\author{F.~Lalibert\'e}
\affiliation{Institut quantique, D\'epartement de physique \& RQMP, Universit\'e de Sherbrooke, Sherbrooke, Qu\'ebec J1K 2R1, Canada}

\author{M.~Dion}
\affiliation{Institut quantique, D\'epartement de physique \& RQMP, Universit\'e de Sherbrooke, Sherbrooke, Qu\'ebec J1K 2R1, Canada}

\author{S.~Badoux}
\affiliation{Institut quantique, D\'epartement de physique \& RQMP, Universit\'e de Sherbrooke, Sherbrooke, Qu\'ebec J1K 2R1, Canada}

\author{N.~Doiron-Leyraud}
\affiliation{Institut quantique, D\'epartement de physique \& RQMP, Universit\'e de Sherbrooke, Sherbrooke, Qu\'ebec J1K 2R1, Canada}

\author{W.A.~Phelan}
\affiliation{Department of Chemistry, Johns Hopkins University, Baltimore, Maryland 21218, USA}
\affiliation{Institute for Quantum Matter, Department of Physics and Astronomy, Johns Hopkins University, Baltimore, Maryland 21218, USA}%

\author{S.M.~Koohpayeh}
\affiliation{Institute for Quantum Matter, Department of Physics and Astronomy, Johns Hopkins University, Baltimore, Maryland 21218, USA}%

\author{W.T.~Fuhrman}
\affiliation{Institute for Quantum Matter, Department of Physics and Astronomy, Johns Hopkins University, Baltimore, Maryland 21218, USA}

\author{J.R.~Chamorro}
\affiliation{Department of Chemistry, Johns Hopkins University, Baltimore, Maryland 21218, USA}
\affiliation{Institute for Quantum Matter, Department of Physics and Astronomy, Johns Hopkins University, Baltimore, Maryland 21218, USA}

\author{T.M.~McQueen}
\affiliation{Department of Chemistry, Johns Hopkins University, Baltimore, Maryland 21218, USA}
\affiliation{Institute for Quantum Matter, Department of Physics and Astronomy, Johns Hopkins University, Baltimore, Maryland 21218, USA}
\affiliation{Department of Materials Science and Engineering, Johns Hopkins
University, Baltimore, MD 21218}%

\author{X.~Wang}
\affiliation{Center for Nanophysics and Advanced Materials, Department of Physics,
University of Maryland, College Park, Maryland 20742, USA}

\author{Y.~Nakajima}
\affiliation{Center for Nanophysics and Advanced Materials, Department of Physics,
University of Maryland, College Park, Maryland 20742, USA}

\author{T.~Metz}
\affiliation{Center for Nanophysics and Advanced Materials, Department of Physics,
University of Maryland, College Park, Maryland 20742, USA}

\author{J.~Paglione}
\affiliation{Center for Nanophysics and Advanced Materials, Department of Physics,
University of Maryland, College Park, Maryland 20742, USA}
\affiliation{Canadian Institute for Advanced Research, Toronto, Ontario M5G 1Z8, Canada}

\author{L.~Taillefer}
\affiliation{Institut quantique, D\'epartement de physique \& RQMP, Universit\'e de Sherbrooke, Sherbrooke, Qu\'ebec J1K 2R1, Canada}
\affiliation{Canadian Institute for Advanced Research, Toronto, Ontario M5G 1Z8, Canada}

\date{\today}

\begin{abstract}
The thermal conductivity $\kappa$ of the Kondo insulator \smb~was
measured at low temperature, down to 70~mK, in magnetic fields up to 15~T,
on single crystals grown using both the floating-zone
{and the flux} method{s}.
The residual linear term $\kappa_0/T$ at $T \to 0$ is found to be zero
in all samples, for all magnetic fields, in agreement with
{previous studies}.
There is therefore no clear evidence of fermionic heat carriers.
In contrast to {some} prior data, we observe a large enhancement
of $\kappa(T)$ with increasing field.
The effect of field is anisotropic, depending on the relative orientation
of field and heat current (parallel or perpendicular), and with respect
to the cubic crystal structure.
We interpret our data in terms of heat transport predominantly by phonons,
which are scattered by magnetic impurities.

\end{abstract}

\pacs{Valid PACS appear here}
\maketitle

\section{INTRODUCTION}

Samarium hexaboride (\smb) is
a Kondo insulator,
%
a material in which the interaction between the localized $f$ electrons
and the conduction band gives rise to a hybridized band structure with a gap~\cite{dzero2010}.
The correlated metallic behavior at high temperature smoothly becomes insulating
below 40~K with the opening of the Kondo gap, giving rise to a diverging
resistance at low temperature.
However, below $\sim 5$~K, a resistivity plateau is observed,
the signature of a metallic state at the surface of the
sample~\cite{xu2013,kim2013,yee2013,zhang2013,wolgast2013,phelan2014}.
This surface state may be topological in nature.

Recently, two independent studies~\cite{li2014,tan2015}
reported the observation of de Haas-van Alphen (dHvA) oscillations in \smb, but with different
interpretations.
In the first, Li~\etal~attributed the quantum oscillations to a two-dimensional Fermi surface associated with
the metallic surface state~\cite{li2014}.
In the second, Tan~\etal~detected additional frequencies
and attributed the quantum oscillations to a three-dimensional Fermi surface associated with
the insulating bulk~\cite{tan2015}.
Although it has been shown that dHvA oscillations can indeed occur in a band insulator
like SmB$_6$~\cite{knolle2015}, a more exotic possibility is the existence of neutral fermions.

%
%

One way to detect mobile fermions
is through their ability to carry entropy,
which a measurement of the thermal conductivity $\kappa(T)$ should in principle detect as a non-zero
residual linear term in the $T=0$ limit, \ie~$\kappa_0/T > 0$.
In this paper, we report low-temperature thermal conductivity measurements in
high-quality single crystals of \smb~down to 70~mK.
In agreement with a prior study~\cite{xu2016}, we observe no residual linear
term $\kappa_0 / T$, at any value of the magnetic field up to 15~T.
However, unlike in that prior study, we observe a large field-induced enhancement
of $\kappa(T)$.
There are two general scenarios for this: either magnetic excitations (like magnons or spinons)
carry heat, or phonons are scattered by a field-dependent mechanism, like magnetic
impurities or spin fluctuations.
In this paper, we argue that our data on \smb~are consistent with the latter scenario,
and discuss in particular the case of phonons scattered by magnetic impurities.

\section{METHODS}

We studied {six} single crystals of \smb, {four grown at Johns
Hopkins University by the floating-zone method (labeled Z1, Z2 and ZC)
and by the flux method (F3), and two grown at University of Maryland
by the flux method (labeled F1 and F2)}.
Growth methods are described elsewhere~\cite{phelan2016,fuhrman2017,PhysRevLett.114.096601,akintola2017}.
\begin{figure*}[!htb]
    \includegraphics[width=0.8\textwidth]{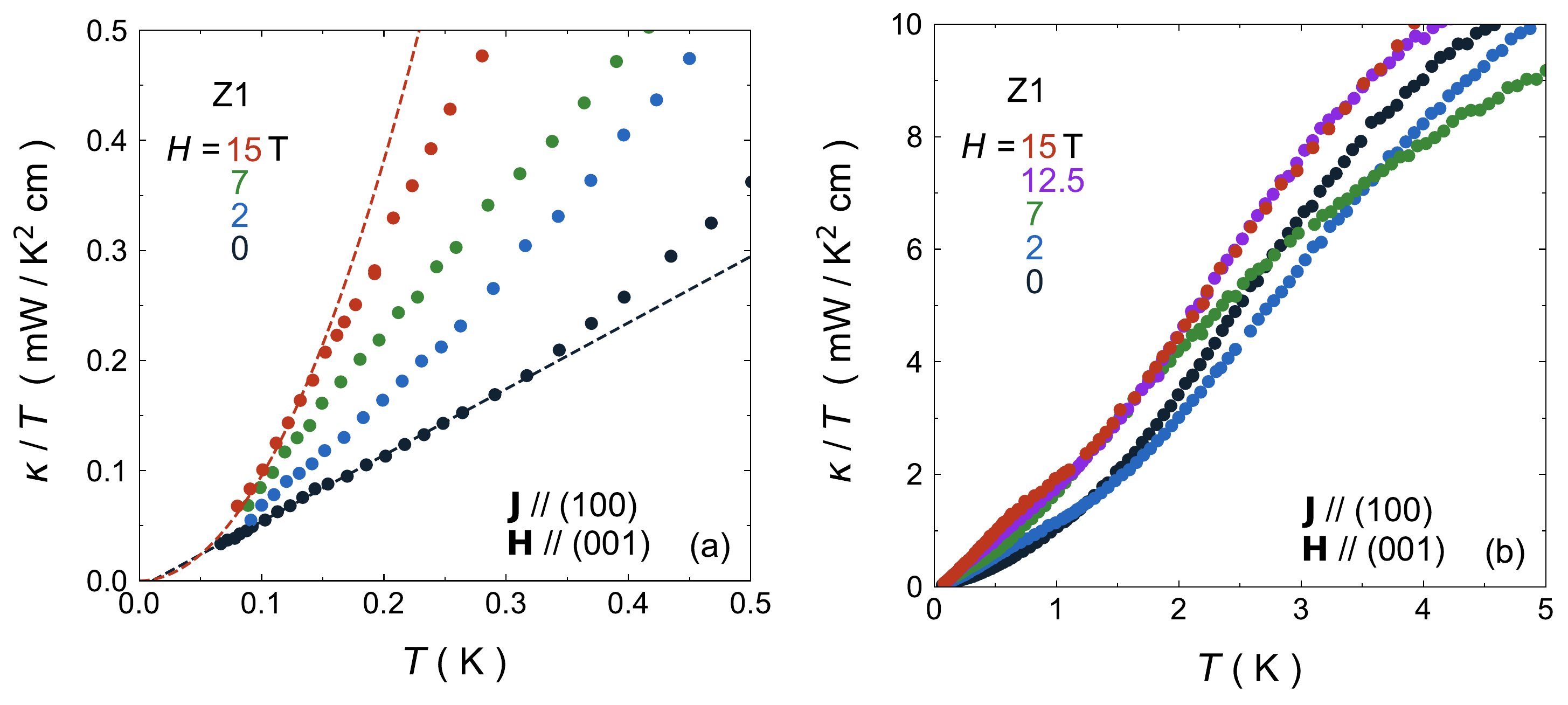}
	\caption{
Thermal conductivity of \smb~plotted as $\kappa/T$ vs $T$ for various values
(as indicated) of the magnetic field applied perpendicular to the heat current,
${\bf H} \parallelsum (001)$ and ${\bf J} \parallelsum (100)$.
(a)
Low temperature regime ($T <0.5$~K).
The black dashed line is a linear fit to the zero-field data below 0.3~K.
The red dashed line is $\kappa = \beta T^3$, with $\beta = 9.5$~mW/K$^4$~cm,
showing that the 15~T data are consistent with $\kappa/T \sim T^2$~below 0.15~K.
Although the field enhances $\kappa(T)$ significantly,
there is no residual linear term at any field, \ie~$\kappa/T \to 0$ as $T \to 0$.
%
(b)
High temperature regime.
The field dependence of $\kappa(T)$ is non-monotonic.
}
\label{thermal_conductivity_low_high_T}
\end{figure*}
%
Sample Z1 was prepared using in-house sources of samarium and boron, while
samples Z2 and ZC
were prepared using commercial sources.
("C" here refers to carbon-doping, in a fraction that is virtually impossible to know
precisely~\cite{phelan2014,phelan2016}, but presumed to be at the 1-5\% level).
The two types of starting material imply different concentrations of rare earth impurities.
The former predominantly contain Gd impurities~\cite{fuhrman2017},
while the latter have more non-magnetic than magnetic impurities.
In addition, samples contain Sm vacancies, typically at the $1$~\% level.
Such vacancies are known to enhance the valence of Sm in \smb~from the non-magnetic Sm$^{2+}$
valence to the $J=5/2$ magnetic Sm$^{3+}$, possibly acting like magnetic impurities~\cite{konovalova1982}.
Surface and bulk magnetic properties of Sm$^{2+}$ and Sm$^{3+}$ were recently studied using
x-ray-absorption spectroscopy and x-ray magnetic circular dichroism
on a sample directly comparable to Z2~\cite{fuhrman2018}, giving insight on the
coupling to magnetic impurities.
%
%
The samples were cut in the shape of rectangular platelets
with the following dimensions (length $\times$ width $\times$ thickness, in \micron) :
$1650 \times 450 \times 300$~(Z1),
$1480 \times 730 \times 218$~(Z2),
$1460 \times 1330 \times 210$~(ZC),
$1375 \times 780 \times 290 $~(F1),
$990 \times 830 \times 218 $~(F2), and 
$2930 \times 830 \times 405 $~(F3).
The contacts were made using H20E silver epoxy.
Room temperature resistivities were 0.23, 0.22, 0.23, 0.18 and 0.16 m$\Omega\>$cm
for Z1, Z2, ZC, F1 and F2, respectively.
Contact resistances were between 100 $\Omega$ and 2 k$\Omega$.
The relative orientation between the crystal structure and the sample dimensions
was determined via x-ray-diffraction measurements primarily using diffraction
peaks (001) and (103).
These measurements also revealed the samples to be good quality single crystals.
The thermal conductivity was measured in a dilution refrigerator down to $70$~mK
with a standard one-heater two-thermometers technique with the heat flowing along
the longest dimension.
The current was injected along the (100) high-symmetry direction
of the cubic crystal structure, \ie~${\bf J}\parallelsum (100)$, {for Z1, F1, F2, and F3,
and along $(110)$ for Z2 and ZC.
%

\begin{figure*}[ht]
    \includegraphics[width=\textwidth]{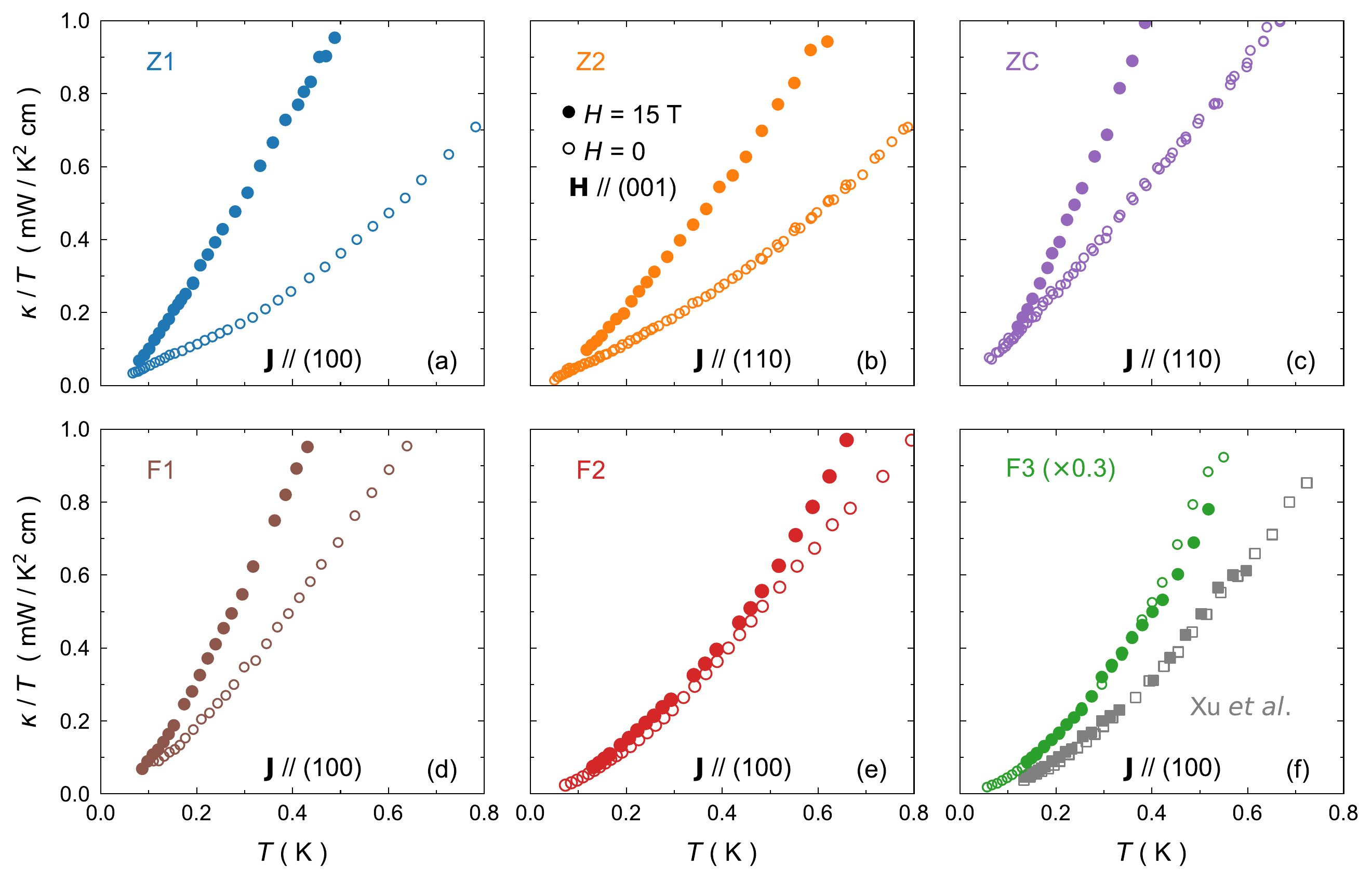}
	\caption{
{
Thermal conductivity of our three zone-grown (upper panels) and flux-grown (lower panels)
samples of \smb, plotted as $\kappa/T$ vs $T$ for $H=0$ (open circles)
and $H=15$~T [full circles; ${\bf H} \parallelsum (001)$].
In panel (f), we reproduce
the data by Xu~\etal~\cite{xu2016} taken on a flux-grown sample ($H=0$, open squares;
$H=14.5$~T, full squares) and compare them with our own flux-grown sample F3 (multiplied by 0.3).}
%
}
\label{compare}
\end{figure*}

%
%

Note that the conductivity of the metallic surface state
makes a completely negligible contribution to the sample's
thermal conductivity, so the measured $\kappa(T)$ is strictly
a property of the insulating bulk.
Indeed, given that the low-$T$ resistivity plateau ranges between $10^{-2}$ and $10^{0}$~$\Omega$~cm
for our samples, and using the Wiedemann-Franz law $\kappa_0/T = L_0 / \rho_0$ where
$L_0 = 2.44\times 10^{-8}$~W~$\Omega$~${\rm K}^{-2}$, we expect a contribution from
the metallic surface state between $10^{-3}$ and $10^{-5}$~mW$/{\rm K}^2$~cm.

\section{RESULTS}

In Fig.~\ref{thermal_conductivity_low_high_T}, the thermal conductivity of sample Z1
is plotted as $\kappa/T$ versus $T$, for various values of the field applied
perpendicular to the heat current ($\bf H \perp \bf J$).
At $T > 2$~K, the effect of the field is non-monotonic:
$\kappa(T)$ decreases with $H$ at first, and then increases.
At $T <0.5$~K (Fig.~\ref{thermal_conductivity_low_high_T}a), the behavior is simpler:
the magnetic field enhances $\kappa(T)$.
This is true for our three zone-grown samples, as shown in Fig.~\ref{compare}.
{A comparable effect is observed for our flux-grown sample F1,
but the enhancement is much smaller for F2 and negligible for F3.
This last result is similar to the previous study by Xu \etal~\cite{xu2016},
performed on a flux-grown crystal (see Fig.~\ref{compare}f).
This sample dependence suggests that an extrinsic mechanism is responsible for the field enhancement.
%
}

To examine whether part of the heat transport in \smb~is carried by fermionic
particles, we first look for a residual linear term.
Simple extrapolation of the data in Fig.~1a yields $\kappa_0/T = 0$ at all fields.
The same is true for all samples (Fig.~2), as also found by Xu \etal~\cite{xu2016}.
Indeed, a linear fit to $\kappa/T$ vs $T$ at $H=0$ describes the data well below 0.3~K,
but it yields a negative value for $\kappa_0/T$.
This means that $\kappa/T$ must go over to a higher power of $T$ at very low $T$,
as expected for phonon conduction, which must go as $\kappa_{\rm p}/T \sim T^2$ in the limit $T \to 0$.
Moreover, the large enhancement with field does not generate a residual linear term.
Indeed, the data at $H=15$~T also extrapolate to zero as $T \to 0$,
and are consistent with $\kappa/T \sim T^2$ below 0.15~K.
Because $\kappa_0/T = 0$ in all samples at all fields,
we are left with no direct evidence of fermionic carriers of heat.
Note, however, that we cannot entirely exclude them, as they could be
present but thermally decoupled from the phonons that bring the heat into the sample~\cite{Smith2005}.

%
%

In Fig.~\ref{thermal_conductivity}, the low-temperature thermal conductivity
at $H=15$~T is shown for different field directions with respect to the crystal
structure axes.
At the lowest temperatures, the effect of a field is isotropic.
However, above a certain temperature (between 0.2 and 0.4~K), an anisotropy develops.
The anisotropy was observed in all floating-zone samples, but only in flux-grown sample F1.
For floating-zone samples, in all cases, the largest conductivity is achieved when
the field is aligned with (110).
For Z2 and ZC, this occurs when the field is applied parallel to the heat current,
but for Z1, (110) is at 45$^\circ$ with respect to the heat current.
For both Z1 and Z2, the conductivity is smallest when the field is along (100),
regardless of the heat current direction.
On the other hand, the anisotropy of sample F1 is reversed, \ie~the conductivity
is largest when the field is along (100), which is also the heat current direction
in this case.
Globally, considering all four samples where an anisotropy is observed, there
seems to be no systematic tendency.
However, among floating-zone samples, the behavior is the same with respect to the
crystal axes.
%
%
%
%

\begin{figure*}[ht]
\centering
\includegraphics[width=0.75\textwidth]{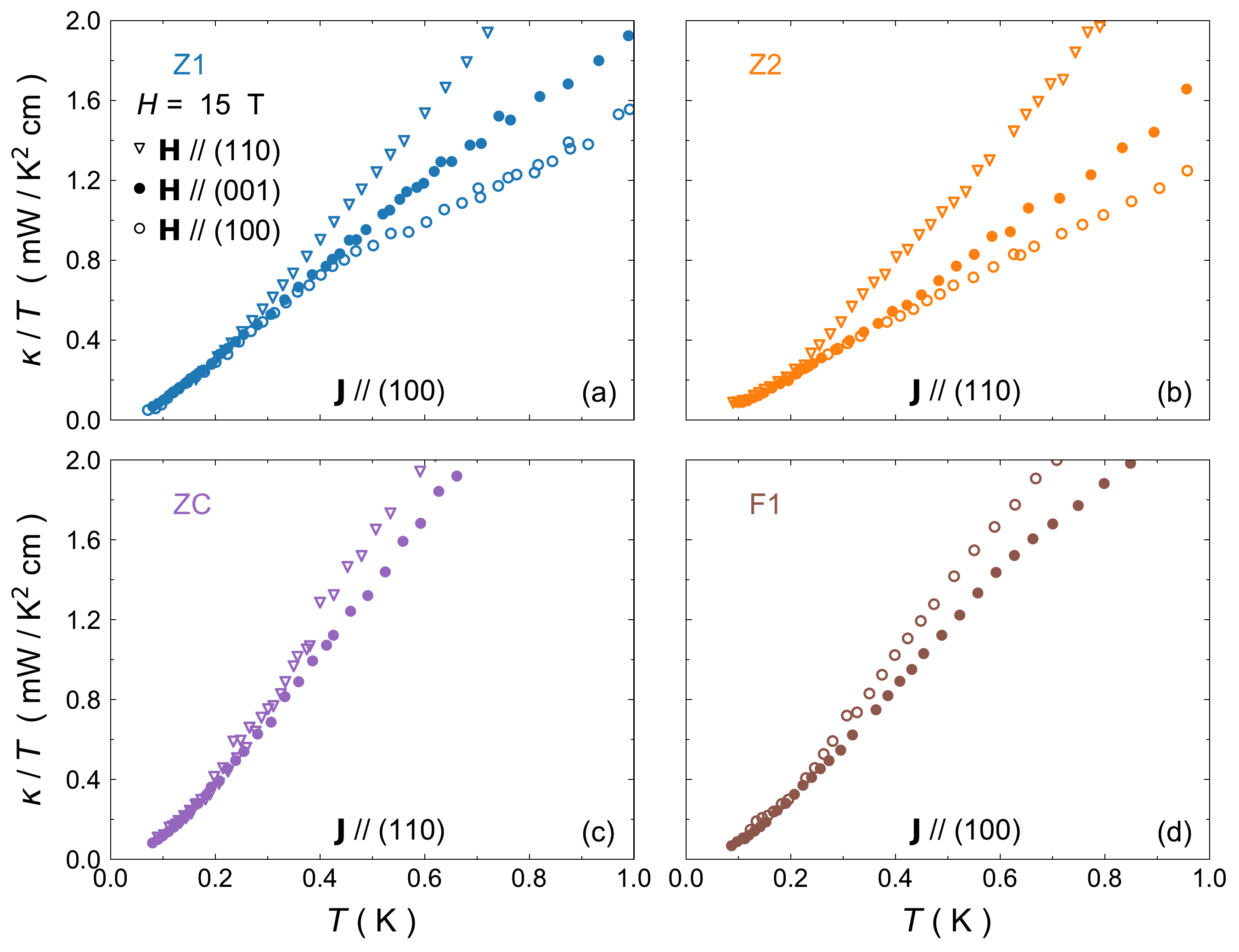}
\caption{
{
Anisotropy of thermal conductivity at $H=15$~T, for a field applied
in different directions (as indicated) with respect to the crystal structure axes.
}
Note that the data are isotropic at the lowest temperatures and
no clear pattern of anisotropy emerges. 
%
}
\label{thermal_conductivity}
\end{figure*}

\begin{figure}
\centering
    \includegraphics[width=0.4\textwidth]{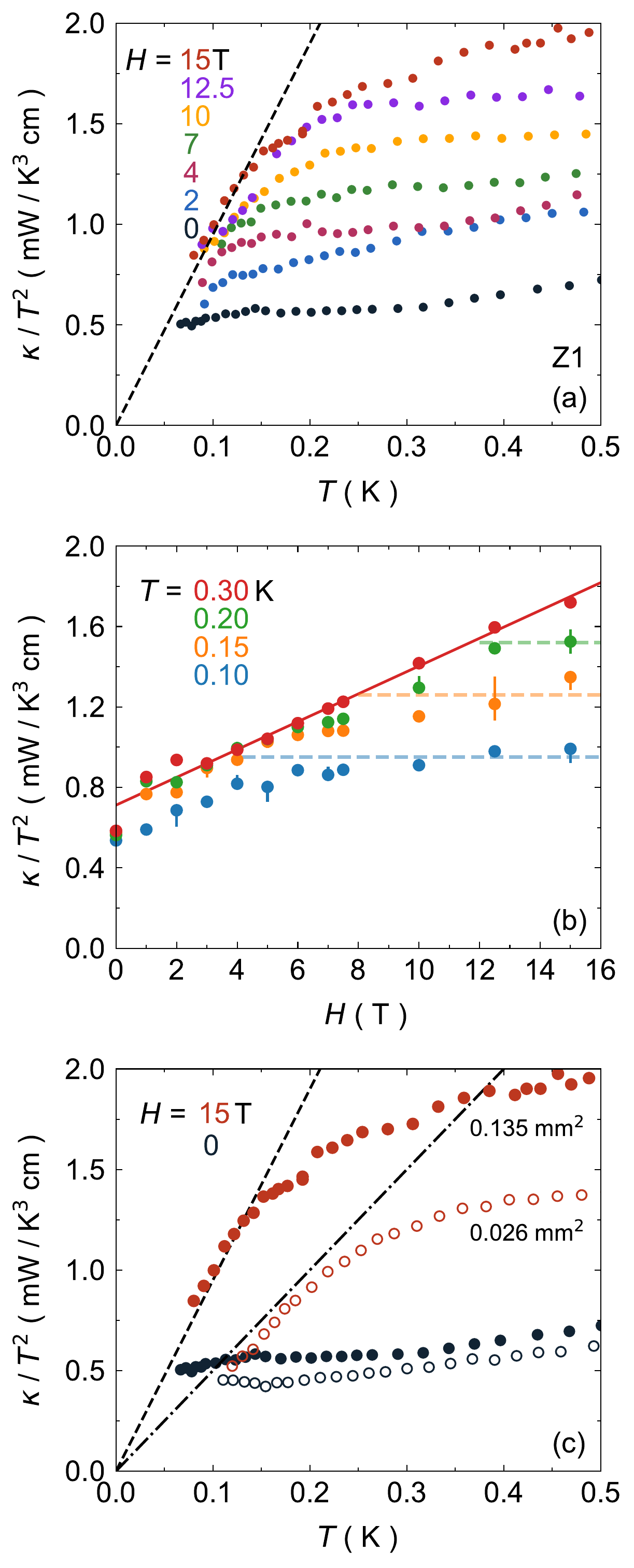}
\caption{
(a)
Same data as in Fig.~\ref{thermal_conductivity_low_high_T}a,
plotted as $\kappa / T^2$ vs $T$.
The black dashed line is $\kappa = \beta T^3$, with $\beta = 9.5$~mW/K$^4$~cm.
(b)
Same data, plotted as $\kappa/T^2$ vs $H$, for four temperatures as indicated.
All lines are guides to the eye.
The red line shows how the data at 0.3~K rise linearly up to the highest field.
The blue line shows how the data at 0.1~K saturate above $H \simeq 6$~T, to a value
consistent with $\beta = 9.5$~mW/K$^4$~cm.
{(c)
Same data at 0 and 15~T from panel (a), compared to the data obtained after polishing
the Z1 sample, reducing the cross-sectional area (width and thickness) by a factor 3.7 (open circles).
The dotted-dashed line is the same as the dashed line, but this slope is reduced by
a factor $\sqrt{3.7}$, as expected from Eq.~\eqref{kappa_ph}.
}
}
\label{balistique}
\end{figure}

\section{DISCUSSION}

Here we discuss a possible scenario for the complex behavior of heat transport
in \smb, where we take the conservative view that the heat is carried entirely by phonons.
%
%
The question is what scatters those phonons.
In the absence of electrons, since \smb~is a bulk insulator, there are two kinds of scattering processes.
The first kind is independent of magnetic field.
It includes sample boundaries, dislocations, grain boundaries, vacancies, and non-magnetic impurities.
Now because $\kappa$ in \smb~is strongly field dependent, there must be a second kind
of scattering process, which depends on field,
involving either low-energy magnetic excitations, such as magnons, or magnetic impurities.
Recent experiments suggest the existence of intrinsic sub-gap excitations,
in the form of either bosonic excitations as observed via inelastic neutron
scattering~\cite{Fuhrman2015} or persistent spin dynamics that extend to very
low temperatures as measured in muon spin relaxation experiments~\cite{akintola2017}.

Because SmB$_6$ samples are known to contain significant levels of rare-earth
impurities and Sm vacancies, phonons are certainly scattered by those impurities.
Even at the 1~\% level,
magnetic impurities can cause a major suppression of the phonon thermal conductivity in insulators
at low temperature~\cite{slack1964}.
%
%
Phonons will scatter most strongly when their energy matches
the difference between the atomic energy levels of the impurity.
Applying a magnetic field will split some of those energy levels.
Increasing the field can therefore make phonons at low $T$ less and less scattered.
This is our proposal for why $\kappa$ in \smb~increases with field at low $T$.
%
%
%

With decreasing temperature, when the phonon mean free path grows to reach the sample dimensions,
it becomes constant and equal to $\ell_0 = 2\sqrt{A/\pi}$, where $A$ is the cross-sectional area of the sample.
(This is strictly true only for scattering off rough surfaces.
Smooth surfaces can cause specular reflection of phonons, yielding a
temperature-dependent (wavelength-dependent)
mean free path that exceeds the sample dimensions~\cite{li2008}.)

\begin{figure*}
\centering
\includegraphics[width=\textwidth]{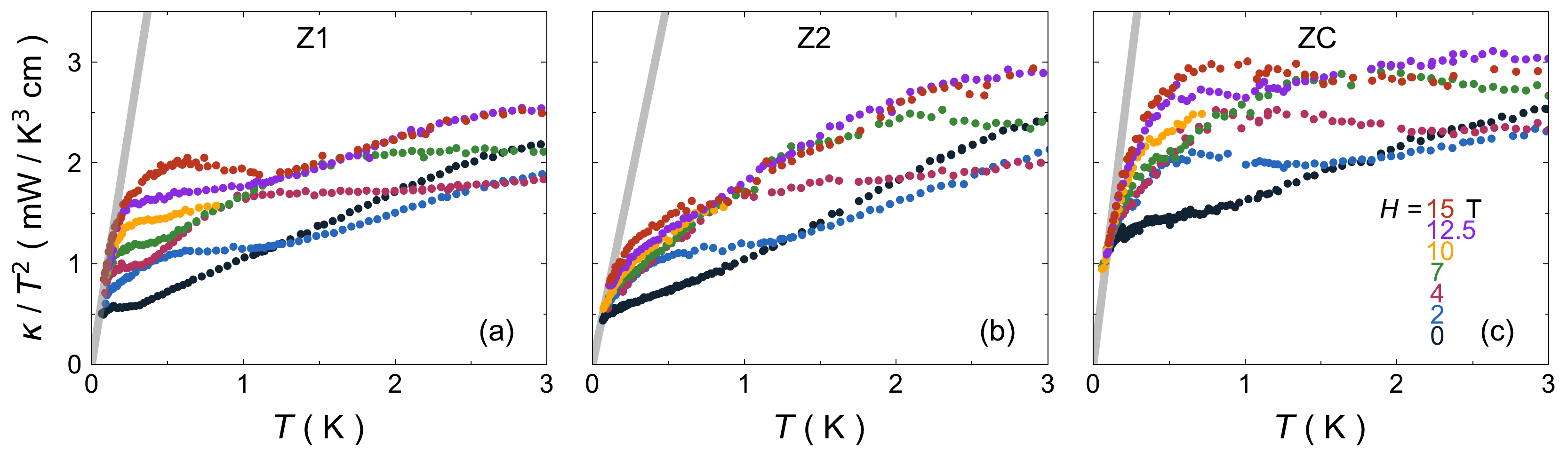}
\caption{
Thermal conductivity of samples Z1 (a), Z2 (b), and ZC (c) plotted as $\kappa/T^2$ vs $T$
up to 3 K and for various magnetic field applied perpendicular to the heat current.
The gray line in each panel shows the estimated boundary-limited phonon contribution,
calculated using Eq.~\eqref{kappa_ph} and the appropriate sample dimensions.}
\label{balistique_high_T}
\end{figure*}
%
In that regime, the phonon thermal conductivity is given by~\cite{li2008}:
\begin{align}
	\kappa_{\rm p}(T) = \frac{2}{15}\pi^2 \kb \qty(\frac{\kb T}{\hbar})^3 v_{\rm p}^{-2} \ell_0
    \label{kappa_ph}
\end{align}
where $v_{\rm p}^{-2}$ is the inverse square of the sound velocity averaged over
three acoustic branches in all $\vb{q}$ directions.

An estimate of the appropriate mean sound velocity may be obtained in terms of the longitudinal ($v_{\rm L}$)
and transverse sound velocities ($v_{\rm T1}$ and $v_{\rm T2}$)~\cite{li2008} :
\begin{align}
    \frac{3}{v_{\rm p}^2} = \frac{1}{v_{\rm L}^2} + \frac{1}{v_{\rm T1}^2} + \frac{1}{v_{\rm T2}^2}~.
\end{align}
Using the elastic constants of \smb~measured at low temperature~\cite{nakamura1991},
we have $v_{\rm L} = 7350 \>{\rm m/s}$, $v_{\rm T1}= 3580 \>{\rm m/s}$ and
$v_{\rm T2}= 6670 \>{\rm m/s}$, giving $v_{\rm p} \simeq 5000$~m/s.
Similar values of $v_{\rm p}$ can be obtained using the Debye temperature
$\Theta_{\rm D}=373$~K~\cite{smith1985} or other estimates~\cite{popov2007}.

Given the dimensions of sample Z1, $\ell_0 = 0.42$~mm,
our estimate of the boundary-limited phonon conductivity is
$\kappa_{\rm p}(T) = \beta T^{3}$ with $\beta = 7 \pm 1$~${\rm mW / K^4 cm}$,
where the error bar reflects the uncertainty on $v_{\rm p}$ and on sample dimensions.
In Fig.~\ref{thermal_conductivity_low_high_T}a and~\ref{balistique}a,
we see that the data for sample Z1 at $H=15$~T are consistent with $\kappa = \beta T^3$ below 0.15~K,
with $\beta \simeq 9$~${\rm mW / K^4 cm}$, a value close to our estimate from Eq.~\eqref{kappa_ph}.
At $H=0$, however, $\kappa(T)$ is well below that.
Our hypothesis is that a magnetic scattering process present in zero field lowers $\kappa$ in \smb,
and this process is quenched or gapped by a field, until it is essentially inactive at $H > 15$~T.

To explore this further, we plot the data as $\kappa/T^2$ versus $T$ in Fig.~\ref{balistique}a.
%
We see that all curves lie below an upper bound given by the straight line
$\kappa/T^2 = \beta T$ with $\beta \simeq 9$~${\rm mW / K^4 cm}$.
As seen from the data plotted as $\kappa/T^2$ versus $H$ in Fig.~\ref{balistique}b,
at $T=0.1$~K the field increases $\kappa$ until it reaches that line, at $H \simeq 6$~T,
above which it saturates.
At $T=0.2$~K, saturation occurs above $H \simeq 12$~T (Fig.~4b).
This behavior is consistent with a scattering process that is gapped by the field.

{
To test this interpretation, we reduced the cross-sectional area of the Z1 sample by
a factor 3.7 (the width was reduced from 450 to 315 \micron, and the thickness was reduced from
300 to 115 \micron), implying that the boundary-limited phonon conductivity is
smaller by a factor $\sqrt{3.7}$.
This is indeed what is observed in Fig.~\ref{balistique}c, confirming our conclusion
that the low $T$ thermal conductivity is set by the boundary limit once the
applied magnetic field has gapped the scattering mechanism present in zero field.
}

\begin{figure*}[!]
\centering
\includegraphics[width=0.85\textwidth]{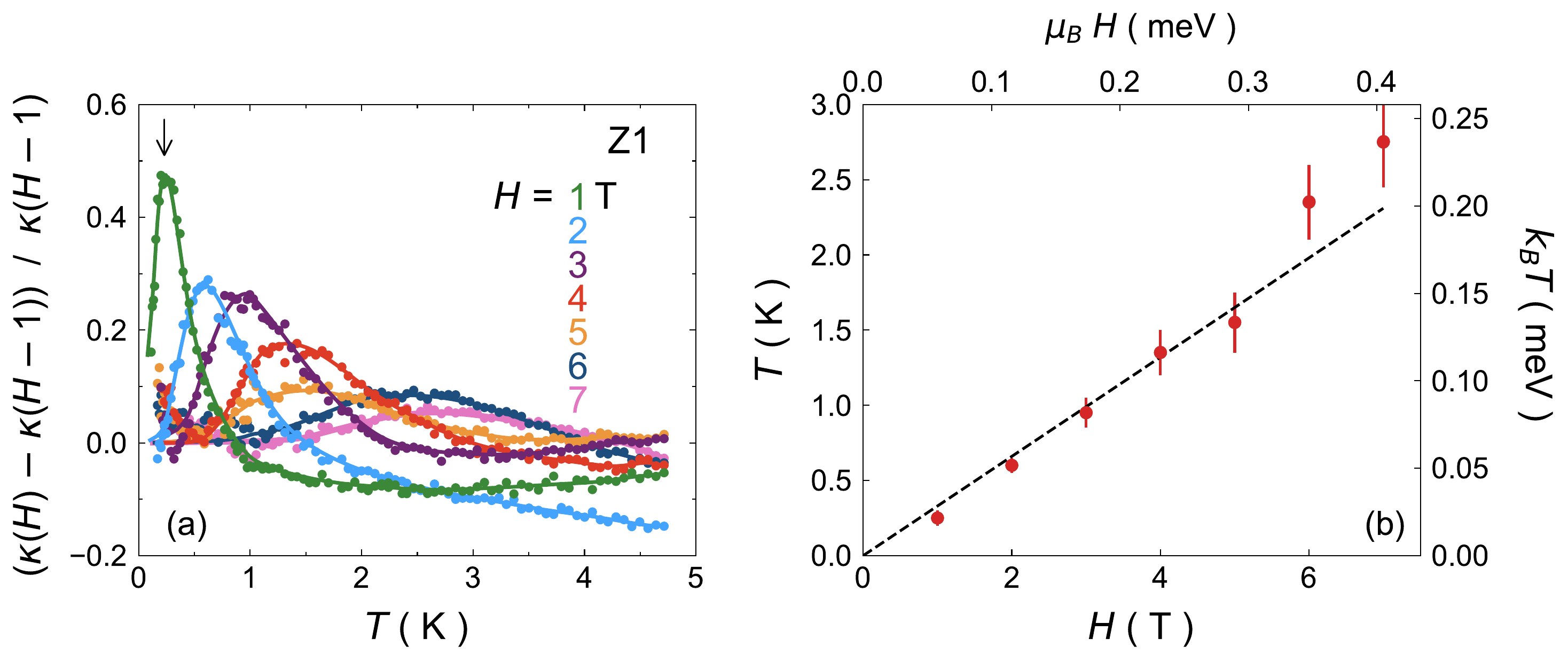}
\caption{(a) Normalized difference between two temperature dependencies for the Z1 sample
measured at two consecutive magnetic fields, with field applied perpendicular to
the heat current.
A maximum in the difference (\eg~vertical arrow for 1~T) is observed at a temperature
that scales with magnetic field.
(b) The temperature of that maximum as a function of magnetic field.
All lines are guides to the eye.}
\label{difference}
\end{figure*}

As seen in Fig.~\ref{balistique_high_T}, the same situation is observed in our three floating zone samples,
whereby $\kappa(T)$ at low $T$ is confined below $\kappa = \beta T^3$,
where $\beta$ is consistent with the value estimated from Eq.~\ref{kappa_ph} given
the particular sample dimensions.
%
%
%
%


We have focused so far on the low-temperature regime below 0.3~K or so.
At higher temperatures, there is a strong but complex field dependence of $\kappa(T)$,
as seen in Figs.~\ref{thermal_conductivity_low_high_T}b and~\ref{balistique_high_T}.
In order to make sense of it, it is instructive to look at the incremental effect of
increasing the field by 1~T, as a function of temperature up to 5~K.
This is shown in Fig.~\ref{difference}a.
The difference between $\kappa(H)$ and $\kappa(H-1)$ reveals a peak at some temperature.
As we increase the field, the peak moves up to higher and higher temperature.
In Fig.~\ref{difference}b, the peak position is seen to scale with $H$ in a roughly linear manner.
This is again consistent with a scenario of magnetic scattering being gapped by the field,
with the gap growing with $H$.
Note, however, that we are now dealing with phonons of much higher energy than before.

It is interesting to look at the energy scales.
Given that the phonons which dominate the thermal conductivity have an energy $E_{\rm ph} \sim 4 k_{\rm B} T$,
Fig.~\ref{difference}b tells us that those phonons are scattered by some mechanism, the characteristic
energy of which is $\Delta \simeq 2 \mu_{\rm B} H$.
If the scattering is associated with the Zeeman splitting of atomic levels,
so that $\Delta = 2 g J \mu_{\rm B} H$,  we get $g J \simeq 1$ from the condition $E_{\rm ph} = \Delta$,
close to the value of $gJ$ for Sm$^{3+}$ vacancies.

{
One of the most puzzling features of the field dependence is its anisotropy relative
to the cubic crystal structure axes.
For floating-zone samples, $\kappa$ is largest for the field parallel to $(110)$,
intermediate for $(001)$, and smallest for $(100)$.
This is true whether the heat current is along $(100)$ or along $(110)$.
For flux sample F1, the anisotropy is reversed: largest for $(100)$ and smallest for $(001)$.
It is difficult to explain the anisotropy given the lack of uniform trend.
One possibility is that the scattering of phonons by magnetic impurities changes
as the field direction changes.
In this scenario, the variation in composition between floating-zone and flux-grown
samples could explain the opposite behavior of F1~\cite{phelan2016,fuhrman2017}.
A theoretical model is needed to understand our observation, which is beyond the
scope of this paper.
%
%
}
%
%
Note that below a certain temperature -- 0.4~K
{for one field orientation and 0.2~K for the other
}
--
there is no anisotropy at $H = 15$~T (Fig.~\ref{thermal_conductivity}).
This is consistent with our interpretation that once the field has gapped the magnetic scattering process,
the phonon mean free path becomes independent of field as it is limited by the sample boundaries,
irrespective of
{
the field orientation.
}

\begin{figure}[h]
\centering
\includegraphics[width=0.4\textwidth]{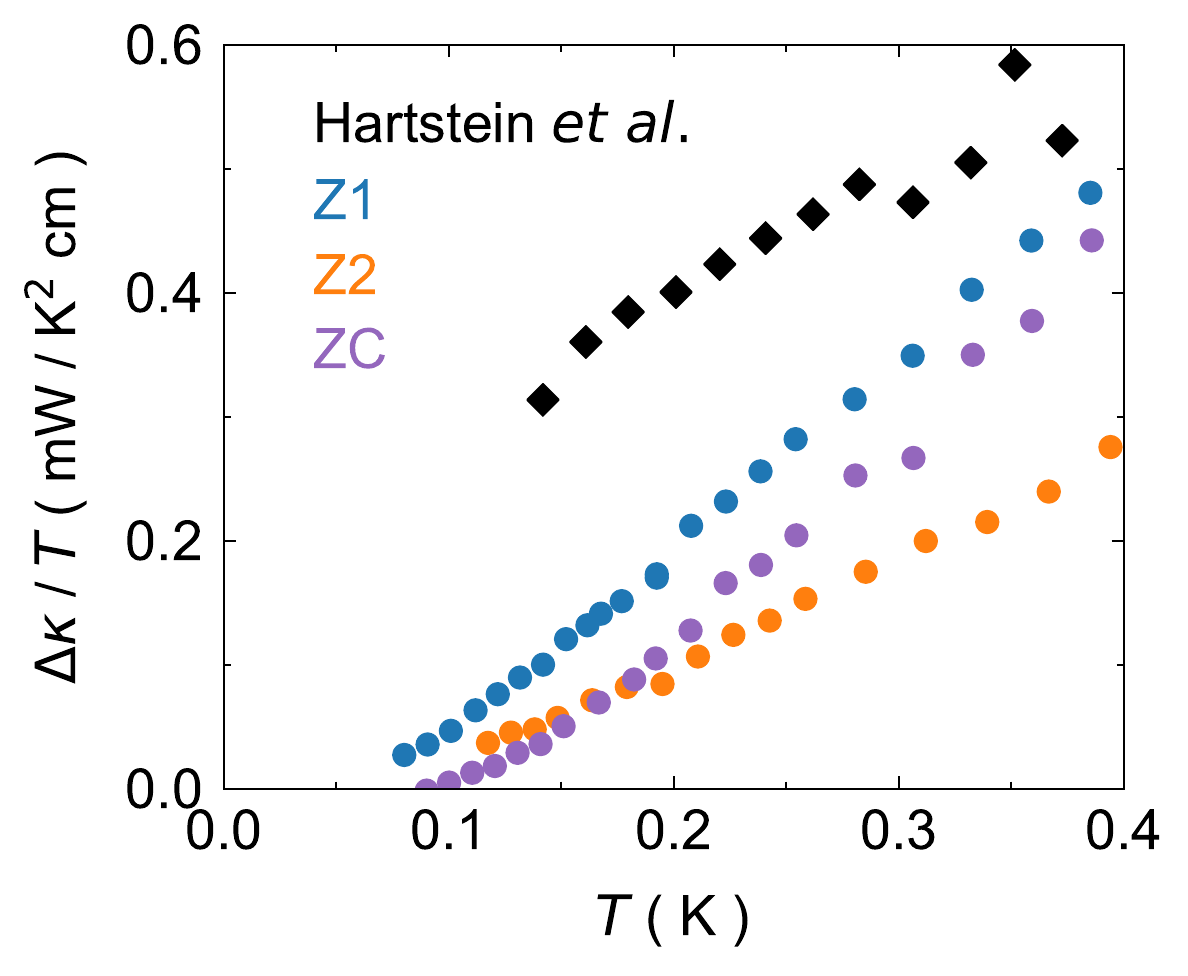}
\caption{
{
%
%
Field-induced thermal conductivity enhancement at low $T$ for the zone-grown samples,
plotted as $\Delta\kappa/T = [\kappa(H_{\rm max}) - \kappa(0)]/T$ vs $T$, where
$H_{\rm max}=15$~T for Z1, Z2 and ZC, and 12~T for the sample of ref.~\cite{hartstein2017} (black diamonds).
}}
\label{cambridge}
\end{figure}

In summary, we propose that rare-earth impurities in \smb,
including Sm vacancies, known to be present
at significant levels in even the best samples~\cite{fuhrman2017},
scatter phonons, and this scattering process is suppressed by a magnetic field, at low temperature.
Theoretical calculations are needed to understand the temperature and field dependence,
as well as the anisotropy of the process.

{
Lastly, we discuss a recent study by Hartstein \etal~on floating-zone grown
samples~\cite{hartstein2017}.
They report a substantial field-induced enhancement of $\kappa(T)$ (well below the
charge gap energy scale) which they interpret as evidence for a Fermi surface
in the bulk arising from novel itinerant low-energy excitations.
This is a very different interpretation to the one we propose here.
%
%
In zero field, their data at low $T$ are well described by Eq.~\ref{kappa_ph}, \ie~$\kappa/T$
goes as $T^2$ with an amplitude expected from their sample's dimensions.
%
Applying a field then increases $\kappa(T)$ beyond the phonon amplitude predicted
by the boundary limit, which they interpret as a fermionic contribution.
Note however that having a purely phononic $\kappa(T)$ larger than the value
predicted byEeq.~\ref{kappa_ph} is not impossible.
Indeed, the boundary limit applies when phonons reflections on the surfaces are diffusive.
If reflections are specular, the mean free path is not limited by the sample's boundaries,
which makes the temperature dependence of $\kappa(T)$ smaller than $T^3$,
and its amplitude larger than in the diffusive case~\cite{li2008}.
}

{
In comparison, our zero-field data are well below the boundary limit.
This reveals that their sample is of higher quality, implying longer phonon mean free paths.
But despite the variation in sample quality, the effect of magnetic field is comparable.
In Fig.~\ref{cambridge}, we plot the difference between $\kappa(T)/T$ at 15~T
and at zero field for our three floating-zone samples, along with the same difference
for their sample (with 12~T).
The enhancement in field is comparable in the sense that it grows smoothly as a
function of temperature, but apparently with an offset of about 0.2~mW$/{\rm K}^2$~cm
in their data.
In contrast to their data, it is unambiguous from our data that $\Delta\kappa/T$ extrapolates
to zero as $T\to0$, in agreement with our conclusion that no residual linear term is
observed at any field.
}

{
If the enhancement in thermal conductivity were really caused by field-induced low-energy
fermionic excitations, there should be a residual linear term.
By measuring at much lower temperature, we confirm that $\kappa_0/T\to0$ even in 15~T.
Moreover, such speculated fermionic excitations would lead to a rapid downturn at
low temperature due to phonon decoupling.
Since neither of these two signatures is observed, we conclude that there is no
experimental evidence of bulk Fermi surfaces in the thermal conductivity of \smb.
}

\vspace{30pt}

\section{SUMMARY}

We have measured the thermal conductivity of
\smb~down to 70~mK, in three zone-grown
and three flux-grown crystals.
No residual linear term was observed in any sample,
either in zero field or in any field up to 15~T.
This means that there is no concrete evidence of fermionic heat carriers in \smb.
However, the field produces a significant enhancement of $\kappa(T)$ in {most} samples.

We interpret our data in a scenario where phonons are the only carriers of heat,
and they are scattered by a magnetic mechanism that is gapped by the field,
such that by 15~T the phonon mean free path grows to reach the sample boundaries
at the lowest temperatures.
The fact that the effect of field depends on its orientation relative to the
crystal structure points to an extrinsic mechanism.
We propose that phonons are scattered by magnetic rare-earth impurities or vacancies.
The fact that the field-induced enhancement of $\kappa(T,H)$ shifts linearly
to higher $T$ with increasing $H$ is consistent with the Zeeman splitting of atomic levels
responsible for the impurity scattering.
%


\section*{ACKNOWLEDGMENTS}

{We acknowledge valuable discussions with P.~Coleman, J.~Knolle, S.~Y.~Li,
J.-Ph.~Reid, M.~Sutherland and W.~Toews.}
We thank S.~Fortier for his assistance with the experiments.
L.T. acknowledges support from the Canadian Institute for Advanced Research and funding from
the Institut Quantique,
the Natural Sciences and Engineering Research Council of Canada (Grant No, PIN:123817),
the Fonds de Recherche du Qu\'{e}bec - Nature et Technologies,
the Canada Foundation for Innovation,
and a Canada Research Chair.
Research at the University of Maryland was supported by AFOSR through Grant
No. FA9550-14-1-0332 and the Gordon and Betty Moore Foundation’s EPiQS
Initiative through Grant No. GBMF4419.
Work at the Institute for Quantum Matter was
supported by the U.S. Department of Energy, Office of Basic
Energy Sciences, Division of Materials Sciences and Engineering
through Grant No. DE-FG02-08ER46544. Partial funding
for this work was provided by the Johns Hopkins University
Catalyst Fund.

\vfill

\bibliography{reference}

\end{document}